\documentclass[%
reprint,
superscriptaddress,
amsmath,amssymb,
aps,
prb,
]{revtex4-2}

\usepackage{graphicx}
\usepackage[pdftex]{color} 
\usepackage{dcolumn}
\usepackage{bm}
\usepackage{here}
\usepackage[utf8]{inputenc}
\usepackage{color}
\usepackage[pdftex]{hyperref}
\hypersetup{
colorlinks=true,
linkcolor=blue,
citecolor=blue,
urlcolor=blue,
setpagesize=false
}

\newcommand{\nn}{\nonumber \\}
\newcommand{\vk}{\bm{k}}
\newcommand{\vq}{\bm{q}}
\newcommand{\ave}[1]{\langle #1 \rangle}
\newcommand{\sk}{\sum_{\bm{k}}} 
\newcommand{\sks}{\sum_{\bm{k} s}} 
\newcommand{\skss}{\sum_{\bm{k} s s^\prime}} 
\newcommand{\cc}[1]{c^\dag_{ #1 }} 
\newcommand{\ca}[1]{c_{ #1 }} 
\newcommand{\qq}{\bm{q} \mathalpha{+} \bm{q}^\prime}

\usepackage{braket}

\begin{document}


\title{Superconducting piezoelectric effect}

\author{Michiya Chazono}
\email[]{chazono.michiya.84s@st.kyoto-u.ac.jp}
\affiliation{Department of Physics, Kyoto University, Kyoto 606-8502, Japan}

\author{Hikaru Watanabe}
\affiliation{RIKEN Center for Emergent Matter Science (CEMS), Wako, Saitama 351-0198, Japan}

\author{Youichi Yanase}
\affiliation{Department of Physics, Kyoto University, Kyoto 606-8502, Japan}
\affiliation{Institute for Molecular Science, Okazaki 444-8585, Japan}


\date{\today}


\begin{abstract}
The magnetopiezoelectric effect (MPE) is a cross-coupling between an electric current and strain in metals with neither inversion symmetry nor time-reversal symmetry. Unlike the conventional piezoelectric effect, the MPE allows a piezoelectric response in superconductors, as we call the superconducting piezoelectric effect (SCPE). The SCPE may enable a piezoelectric response without Joule heating and provide a probe of exotic superconducting symmetry. In this paper, we propose a formulation of the SCPE and calculate both the MPE and SCPE in the two-dimensional noncentrosymmetric s-wave superconductor under an in-plane magnetic field. We find that the magnitude of the SCPE is comparable to the MPE. It is also clarified that finite total momentum of Cooper pairs in the helical superconducting state plays a crucial role in the SCPE. 
\end{abstract}

\maketitle

\section{INTRODUCTION}
Lack of inversion symmetry (IS) allows various physical responses prohibited in materials with IS, such as the piezoelectric effect (PE)~\cite{Curie1894}, Edelstein effect \cite{Edelstein1990}, natural optical activity \cite{Landau_electromagnetism}, and so on. The IS breaking also influences the quantum phase of matter. For instance, the mixing between the spin-singlet and spin-triplet pairings occurs in noncentrosymmetric superconductors~\cite{Bauer2012}. The strong parity mixing is of interest since it is regarded as a key to the topological superconductivity \cite{Smidman2017}.

In materials with neither IS nor time-reversal symmetry (TRS), a richer variety of phenomena can be realized; \textit{e.g.} the magnetoelectric effect \cite{Fiebig2005,Dong2015} and nonreciprocal response \cite{Tokura2018,Ideue2021}. The nonreciprocal phenomena in superconductors lacking both IS and TRS are reported recently \cite{Wakatsuki2017, Itahashi2020, Ando2020, Nakamura2020, Baumgartner2021} and attracting much attention.

The magnetopiezoelectric effect (MPE), the electric-current-induced lattice distortion, is one of the recently discovered phenomena in materials with neither IS nor TRS. In the linear response regime, the MPE response formula is given by
\begin{align}
s_{ij} = d_{ijk} J_k,
\label{MPE}
\end{align}
where $s_{ij}$ is a strain tensor, $d_{ijk}$ is a MPE coefficient, and $J_k$ is an electric current. This effect resembles the conventional PE, namely, the electric-field-induced lattice distortion,
\begin{align}
s_{ij} = d^\prime_{ijk} E_k.
\label{CPE}
\end{align}
At first glance, it seems that we merely 
replace an electric field $E_k$ with $J_k$ by relating them through an electric conductivity. However, the parities under the time-reversal operation are opposite between $J_k$ and $E_k$, \textit{i.e.}, $-1$ for $J_k$ and $+1$ for $E_k$. Therefore, while the conventional PE does not require TRS breaking, the MPE is realized only in materials lacking both IS and TRS. 

The inverse MPE was predicted as generalization of the magnetoelectric effect in noncentrosymmetric metals under an external magnetic field \cite{Varjas2016}. Another study proposed the MPE in antiferromagnetic metals whose order parameter has the same symmetry properties as the odd-parity magnetic multipole moment~\cite{Watanabe2017}. The IS and TRS symmetries are simultaneously broken in such antiferromagnets. 
Experiments have been performed for the latter with antiferromagnetic metals EuMnBi$_2$ and CaMn$_2$Bi$_2$, and the MPE has been actually observed~\cite{Shiomi2019Mar, Shiomi2019Aug}. Remarkably, the latest experimental result reveals that the MPE response becomes larger with the higher conductivity \cite{Shiomi2020}. This is consistent with the MPE which occurs only in metals~\cite{Watanabe2017}, while it contrasts with the fact that the conventional PE is suppressed by the higher conductivity.
These results point to the solid discovery of the MPE and show that metals are good candidates for lead-free piezoelectric materials. On the other hand, it has also been pointed out that the MPE is largely influenced by the Joule heating \cite{Shiomi2019Aug}. For practical applications as well as for establishing scientific grounds, it is desirable to explore the Joule-heating-free MPE. 

In this paper, we 
propose a piezoelectric response in superconductors. The conventional PE is prohibited in superconductors because the electric field should vanish due to the zero resistance. On the other hand, the Joule-heating-free supercurrent can flow, and the supercurrent-induced lattice distortion similar to the MPE can occur.
We call this phenomenon the superconducting piezoelectric effect (SCPE), which will be one of the proposals for the Joule-heating-free MPE. 
Here we emphasize that the SCPE and the MPE are essentially different phenomena as we will see below. Properties of the PE, MPE, and SCPE are summarized in Table~\ref{table:PEs}.
\begin{table*}[!]
 \caption{Comparison of the PE, MPE, and SCPE. Symmetry requirement, source field, and presence or absence of dissipation are summarized. The PE, MPE, and SCPE occur in insulators and semiconductors, metals, and superconductors, respectively.}
 \label{table:PEs}
 \centering
  \begin{tabular}{cccc}
  \hline
   & PE & MPE & SCPE \\
  \hline \hline
   Symmetry condition & IS breaking & IS and TRS breaking & IS and TRS breaking \\
   Source & Electric field & Normal electric current & Supercurrent \\
   Dissipation & Absent & Present & Absent \\ 
   System & Insulators, Semiconductors & Metals & Superconductors \\
   \hline
  \end{tabular}
\end{table*}
Irrespective of a practical application, the SCPE is expected to be a probe of IS and TRS breaking in superconductors since it is sensitive to symmetry breaking.
Moreover, as we show later, we could utilize the SCPE to uncover the superconducting state in detail. 

As a representative example, we study the SCPE in two-dimensional Rashba $s$-wave superconductors under an in-plane magnetic field. In Sec.~\hyperref[sec:Formulation]{II}, we present the model Hamiltonian and formulate the SCPE. In Sec.~\hyperref[sec:Result]{III}, we compare the numerical results of the MPE and SCPE and find that the obtained SCPE response is comparable to the MPE response. By analyzing the chemical potential and magnetic field dependence, we clarify that the finite total momentum of Cooper pairs in the helical superconducting state plays an essential role in the SCPE. Finally, we summarize our results and discuss a prospect in Sec.~\hyperref[sec:Summary]{IV}.

\section{FORMULATION}
\label{sec:Formulation}
\subsection{Model Hamiltonian}
To demonstrate the SCPE, we study two-dimensional $s$-wave superconductors with $C_{4v}$ crystal structure under an in-plane magnetic field based on the following Bogoliubov-de Gennes (BdG) Hamiltonian
\begin{align}
&H = H_\textrm{kin} + H_\textrm{ASOC} + H_\textrm{Zeeman} + H_\textrm{s-wave},
\label{HSC} \\
&H_{\textrm{kin}} = \sks \varepsilon (\vk) \cc{\vk,s} \ca{\vk,s},
\label{Hkin} \\
&H_{\textrm{ASOC}} = \skss \bm{g}(\vk) \cdot 
\bm{\sigma}_{s s^\prime}  \cc{\vk,s} \ca{\vk,s^\prime},
\label{HASOC} \\
&H_{\textrm{Zeeman}} = - \mu_{\textrm{B}} \skss \bm{H} \cdot 
\bm{\sigma}_{s s^\prime}  \cc{\vk,s} \ca{\vk,s^\prime},
\label{HZeeman} \\
&H_\textrm{$s$-wave} = \sk \left( \Delta_0 \cc{\vk+\vq,\uparrow} \cc{-\vk+\vq,\downarrow} + \textrm{h.c.}  \right),
\label{Hswave}
\end{align}
where $\varepsilon (\vk) = 2 t_1 (\cos{k_x} + \cos{k_y}) + 4 t_2 \cos{k_x} \cos{k_y} - \mu,~ \bm{g}(\vk) = \alpha (\sin{k_y}, - \sin{k_x},0)$, $\bm{H} = (0, H_0,0)$, $\bm{\sigma} = (\sigma_x, \sigma_y, \sigma_z)$ is the vector of Pauli matrices, and $\ca{\vk,s}~(\cc{\vk,s})$ is the annihilation (creation) operator with momentum $\vk$ and spin $s$. $H_\textrm{kin}$ is a kinetic energy in the tight-binding approximation measured from a chemical potential $\mu$, $H_{\textrm{ASOC}}$ is a Rashba-type spin-orbit coupling, $H_{\textrm{Zeeman}}$ is a Zeeman field, and $H_{\textrm{$s$-wave}}$ represents a $s$-wave superconducting order parameter introduced phenomenologically. In the Rashba superconductor under the in-plane magnetic field, the helical superconducting state is realized with a finite total momentum of Cooper pairs $2 \vq = (2 q_0, 0)$ without injecting an electric current \cite{Smidman2017,Bauer2012}. We can rewrite the Hamiltonian in the matrix form by using the Nambu spinor $\bm{c}_{\vk,\vq} = (\ca{\vk+\vq,\uparrow}, \ca{\vk+\vq,\downarrow}, \cc{-\vk+\vq,\uparrow}, \cc{-\vk+\vq,\downarrow})^T$,
\begin{align}  
H 
&= \frac{1}{2} \sk \bm{c}^\dagger_{\vk,\vq} H(\vk,\vq) \bm{c}_{\vk,\vq} \nn
&=
\frac{1}{2} \sk 
\bm{c}^\dagger_{\vk,\vq}
\left(
\begin{array}{cc}
 H_{\textrm{N}}(\vk + \vq) & \Delta_0 (i \sigma_y) \\
 \Delta_0 (i \sigma_y)^T & -H_{\textrm{N}}(-\vk + \vq)^T
\end{array}
\right)
\bm{c}_{\vk,\vq},
\label{Hmat}
\end{align}
where $H_{\textrm{N}}(\vk)$ corresponds to the normal state Hamiltonian
\begin{align}  
& H_{\textrm{N}}(\vk) =
\left(
\begin{array}{cc}
 \varepsilon (\vk) & g_{-}(\vk) \\
 g_{+}(\vk) & \varepsilon (\vk)
\end{array}
\right), \label{HN} 
\end{align}
with $g_{\pm} = g_x(\vk) \pm i(g_y(\vk) - \mu_{\textrm{B}} H_0) $.

\subsection{MPE mode}
\begin{figure}[b]
  \centering
 \includegraphics[width=80mm]{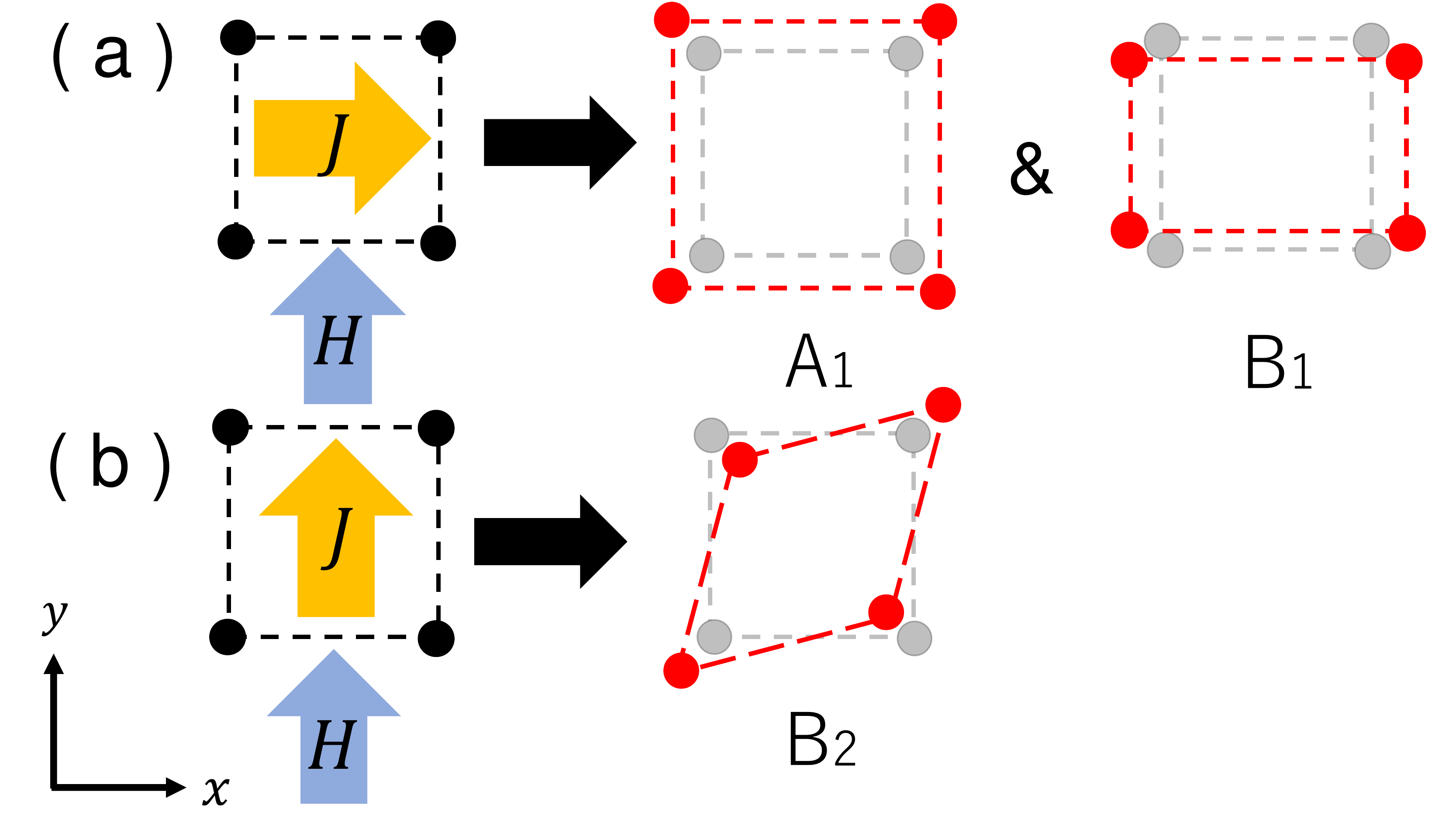}
  \caption{Schematics of the MPE and SCPE. Lattice distortion induced by the electric current is illustrated. 
(a) A$_1$ and B$_1$ modes with the electric current flowing in the $x$-direction. (b) B$_2$ mode with the electric current flowing in the $y$-direction.
  }
\label{fig:nem}
\end{figure}

According to the symmetry argument, 
the three MPE modes are  realizable in this model: the A$_1$ and B$_1$ modes when the electric current flows in the $x$-direction, ${\bm J} \parallel \hat{x}$, and the B$_2$ mode when ${\bm J} \parallel \hat{y}$. 
Figure~\ref{fig:nem} illustrates the three MPE modes.
For other setting different from Fig.~\ref{fig:nem}, the relation between the current direction ($J \perp H$ or $J \parallel H$) and the MPE modes changes, as shown in Appendix A.

To define the MPE mode, we introduce the weighted density operator $\hat{n}_{i} $ for $i=$A$_1$, B$_1$, and B$_2$, 
\begin{align}
&\hat{n}_i = \frac{1}{V} \sks D_i (\vk) \cc{\vk , s} \ca{\vk , s},
\label{nnem} \\
&D_{\textrm{A}_{1}}(\vk) = \cos{k_x} + \cos{k_y}, \nn
&D_{\textrm{B}_{1}}(\vk) = \cos{k_x} - \cos{k_y}, 
\label{nem} \\
&D_{\textrm{B}_{2}}(\vk) = 2 \sin{k_x}  \sin{k_y}. \nonumber 
\end{align}
The operators characterize the modulation of hopping parameters coupled to 
the lattice distortion. The coefficient is determined to make the norms equivalent.
Note that we classify these modes based on the irreducible representations of the $C_{4v}$ point group, which is the symmetry of the system with no magnetic field. Significantly, the B${_1}$ mode belongs to the totally symmetric representation when we consider the symmetry reduction due to the magnetic field. We adopt the classification based on high symmetry 
to distinguish between the quadrupole strain B${_1}$ mode and the expansion-shrink A${_1}$ mode.

\subsection{SCPE and MPE}
We formulate the SCPE and MPE on an equal footing by the coupling between the weighted density and electric current
\begin{align}
&\ave{\hat{n}_{\textrm{A}_1}} = d_{\textrm{A}_1} \ave{\hat{J}_x}, \nn
&\ave{\hat{n}_{\textrm{B}_1}} = d_{\textrm{B}_1} \ave{\hat{J}_x}, 
\label{MPE2} \\
&\ave{\hat{n}_{\textrm{B}_2}} = d_{\textrm{B}_2} \ave{\hat{J}_y}, \nonumber
\end{align}
where $\hat{\bm{J}}$ is the electric current operator and $\ave{\cdots}$ represents the expectation value in the perturbed system.
The lattice strain is obtained by the weighted density via the electron-lattice coupling, $s_i = C_{\rm el}^i \ave{\hat{n}_{i}}$. We leave an evaluation of the material parameter $C_{\rm el}^i$ as a future study and regard $d_{\textrm{A}_1}$, $d_{\textrm{B}_1}$, $d_{\textrm{B}_2}$ as the MPE and SCPE coefficients. 
In this subsection, we specifically consider the case of the A$_1$ mode for an example.
Formulas for the B$_1$ and B$_2$ modes are straightforwardly obtained. 

We define the SCPE as distortion induced by supercurrent, and thus, we can formulate it as an equilibrium phenomenon. When a small supercurrent flows in the $x$-direction, the Cooper pairs get corresponding total momentum $2 \vq^\prime = (2 q^\prime, 0)$ in addition to $2 \vq$ in the static state. Thus, 
the supercurrent-flowing state is described by the Hamiltonian \eqref{HSC} where momentum $\vq$ is replaced with $\qq$ in Eq.~\eqref{Hmat},
\begin{align}  
H
= \frac{1}{2} \sk \bm{c}^\dagger_{\vk,\qq} H(\vk,\qq) \bm{c}_{\vk,\qq}.
\label{Hmat2}
\end{align}
Using the Nambu spinor $\bm{c}_{\vk,\qq}$, the operators $\hat{n}_{\textrm{A}_1}$ and $\hat{J}_x$ are written as
\begin{align}
&\hat{n}_{\textrm{A}_1}
= \sk \bm{c}^\dagger_{\vk,\qq} n_{\textrm{A}_1}(\vk,\qq) \bm{c}_{\vk,\qq}, \label{nA1} \\
&\hat{J}_x
= \sk \bm{c}^\dagger_{\vk,\qq} J_x (\vk,\qq) \bm{c}_{\vk,\qq}, \label{Jx}
\end{align}
where the matrices are given by
\begin{align}
&n_{\textrm{A}_1}(\vk,\qq) = \frac{1}{2}
\scriptsize{
\left(
 \begin{array}{cc} 
 D_{\textrm{A}_1}(\vk \mathalpha{+} \qq) \mathalpha{\times} I_2 & 0 \\
 0 & \mathalpha{-} D_{\textrm{A}_1}(\mathalpha{-} \vk \mathalpha{+} \qq) \mathalpha{\times} I_2 
 \end{array}
\right)
}, \\
&J_x(\vk,\qq) = \frac{e}{2}
\scriptsize{
\left(
 \begin{array}{cc} 
 \partial H_\textrm{N}(\vk+\qq)/\partial k_x & 0 \\
 0 & -\partial H_\textrm{N}(-\vk+\qq)^T/\partial k_x
 \end{array}
\right)
}. 
\end{align}
Therefore, expectation values are calculated by
\begin{align}
\ave{\hat{n}_{\textrm{A}_1}}_{\textrm{eq}, \vq + \vq^\prime} = \frac{1}{V} \sum_{\vk \alpha} 
[\tilde{n}_{\textrm{A}_1}(\vk,\qq) ]_{\alpha \alpha} f ( E_\alpha (\vk, \qq) ), \label{aven} \\
\ave{\hat{J}_x}_{\textrm{eq}, \vq + \vq^\prime} = \frac{1}{V} \sum_{\vk \alpha} 
[\tilde{J}_x (\vk,\qq) ]_{\alpha \alpha} f ( E_\alpha (\vk, \qq) ), \label{aveJ}
\end{align}
where $E_\alpha$ are eigenvalues of the Hamiltonian \eqref{Hmat2}
\begin{align}
E_\alpha (\vk,\qq) = [U^\dag (\vk,\qq) H(\vk,\qq) U(\vk,\qq)]_{\alpha \alpha},
\end{align}
and $f ( E )$ is the Fermi distribution function. We introduced $\tilde{n}_{\textrm{A}_1} = U^\dag n_{\textrm{A}_1} U$ and $\tilde{J}_x = U^\dag J_x U$ for the band representations of $\hat{n}_{\textrm{A}_1}$ and $\hat{J}_x$, respectively. Then, we define the SCPE coefficient for 
the A$_1$ mode by
\begin{align}
d^\textrm{(SC)}_{\textrm{A}_1} = 
\left. \frac{d\ave{\hat{n}_{\textrm{A}_1}}_\textrm{eq}}{d\ave{\hat{J}_x}_\textrm{eq}} \right|_{q^\prime = 0}.
\label{SCPE}
\end{align}
Other SCPE coefficients, $d^\textrm{(SC)}_{\textrm{B}_1}$ and $d^\textrm{(SC)}_{\textrm{B}_2}$,
are defined in the same way.

In contrast to the SCPE, the MPE is induced by a dissipative current, and thus it is a non-equilibrium phenomenon. Therefore, we should calculate it by the linear response theory, and the formulation has been established using the Kubo formula in the previous study \cite{Watanabe2017}. The susceptibility $\chi^\textrm{(N)}_{\textrm{A}_1}$ is given by
\begin{align}
\chi^\textrm{(N)}_{\textrm{A}_1}
&= \frac{\ave{\hat{n}_{\textrm{A}_1} - \ave{\hat{n}_{\textrm{A}_1}}_{\textrm{eq}}}}{E_x},\\
&= \frac{- i e}{V} \sum_{\vk,n,m}
\frac{[\tilde{n}_{\textrm{A}_1}(\vk)]_{nm} [\tilde{v}_x(\vk)]_{mn}}{E_n (\vk) - E_m (\vk) + i \delta}  \frac{f(E_n) - f(E_m)}{E_n (\vk) - E_m (\vk)},
\label{chi}
\end{align} 
where $n$ and $m$ are indices for the eigenstates of the normal Hamiltonian~\eqref{HN}, and $v_x(\vk) = \partial H_{\textrm{N}}(\vk)/\partial k_x$ is the velocity operator in the Bloch representation. Here $\tilde{n}_{\textrm{A}_1}$ and $\tilde{v}_x$ denote the band representation in the normal state, and $\delta$ is the infinitesimal quantity introduced to assume an adiabatic procedure. In our calculation, $\delta$ is regarded as a scattering rate and assumed to be a small finite value (the relaxation time approximation). In the normal state, the matrix in Eq.~\eqref{nnem} is proportional to the identity matrix in the subspace spanned by the spin degree of freedom, and therefore, we can simplify Eq.~\eqref{chi} to 
\begin{align}
\chi^\textrm{(N)}_{\textrm{A}_1} 
= \frac{- e}{V \delta} \sum_{\vk,n}
D_{\textrm{A}_1}(\vk) [\tilde{v}_x(\vk)]_{nn} \frac{\partial f(E_n)}{\partial E}.
\label{chi2}
\end{align}

The Kubo formula, Eq.~\eqref{chi2}, represents the response to the electric field $E_x$ instead of the electric current $J_x$. To define the MPE in the form of current-induced phenomenon as Eq.~\eqref{MPE2}, we rewrite the response formula by calculating the electric conductivity using the Kubo formula 
\begin{align}
\sigma_{x}
= \frac{- e^2}{V\delta} \sum_{\vk,n,m}
[\tilde{v}_x(\vk)]_{nn} [\tilde{v}_x(\vk)]_{nn} \frac{\partial f(E_n)}{\partial E},
\label{sigma}
\end{align} 
and define the MPE coefficient $d^\textrm{(N)}_{\textrm{A}_1}$ by 
\begin{align}
d^\textrm{(N)}_{\textrm{A}_1} 
= \frac{\chi^\textrm{(N)}_{\textrm{A}_1}}{\sigma_{x}}
\left(= \frac{\ave{\hat{n}_{\textrm{A}_1} - \ave{\hat{n}_{\textrm{A}_1}}_\textrm{eq}}}{\ave{\hat{J}_x}} \right).
\label{NMPE}
\end{align}
As mentioned above, we adopt the relaxation time approximation. Thus, the susceptibility $\chi^\textrm{(N)}_{\textrm{A}_1}$ and conductivity $\sigma_{x}$ are depending on the scattering rate $\delta$. However, it is clear from Eq.~\eqref{NMPE} that the MPE coefficient $d^\textrm{(N)}_{\textrm{A}_1}$ is intrinsic because it is independent of $\delta$.

\section{RESULT}
\label{sec:Result}
In this section, we show the numerical results of the MPE and SCPE on the basis of the model Hamiltonian \eqref{HSC}. We take $e = 1$ and $\mu_B = 1$ for simplicity and adopt parameters $t_1 = 1.0,~ t_2 = -0.15,~ \alpha = 0.30,~ \Delta_0 = 0.10$ and the temperature $T =0.01$ unless we explicitly state otherwise. 

\subsection{MPE}
\begin{figure}[t]
  \centering
 \includegraphics[width=80mm]{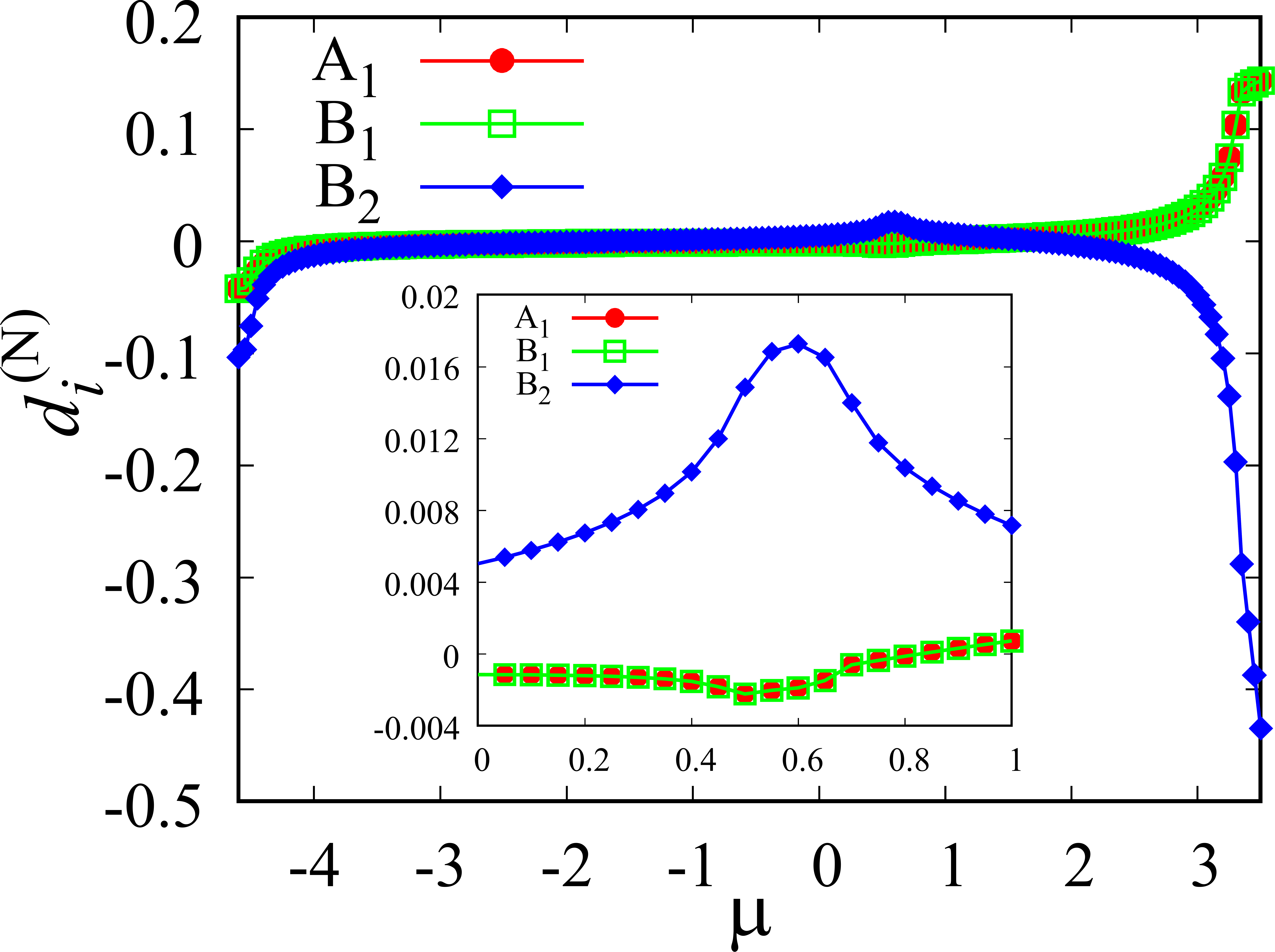}
  \caption{
MPE coefficients 
$d^\textrm{(N)}_{\textrm{A}_1}$,  $d^\textrm{(N)}_{\textrm{B}_1}$ and $d^\textrm{(N)}_{\textrm{B}_2}$ as functions of the chemical potential. 
Note that $d^\textrm{(N)}_{\textrm{A}_1}$ and $d^\textrm{(N)}_{\textrm{B}_1}$ are exactly the same.
The inset shows the region around $\mu = 0.6$. 
 }
\label{fig:nmpe}
\end{figure}

First, we show the results of the MPE. 
The chemical potential dependence of the MPE coefficients $d^\textrm{(N)}_i$ under the magnetic field $H_0 = 0.10$ is shown in Fig.~\ref{fig:nmpe}. 
Here $d^\textrm{(N)}_{\textrm{A}_1}$ and $d^\textrm{(N)}_{\textrm{B}_1}$ are exactly the same, as we prove in Appendix B. Thus, we discuss only the A$_1$ and B$_2$ modes in the rest of this subsection.

We see similar behaviors in
$d^\textrm{(N)}_{\textrm{A}_1}$ and $d^\textrm{(N)}_{\textrm{B}_2}$; 
their magnitudes significantly increase in the low carrier density region and show a small peak around $\mu = 0.6$. To understand these common features, we calculate the band-resolved contributions to the susceptibility $\chi^\textrm{(N)}_{i}$.
\begin{figure}[b]
  \centering
  \includegraphics[width=80mm]{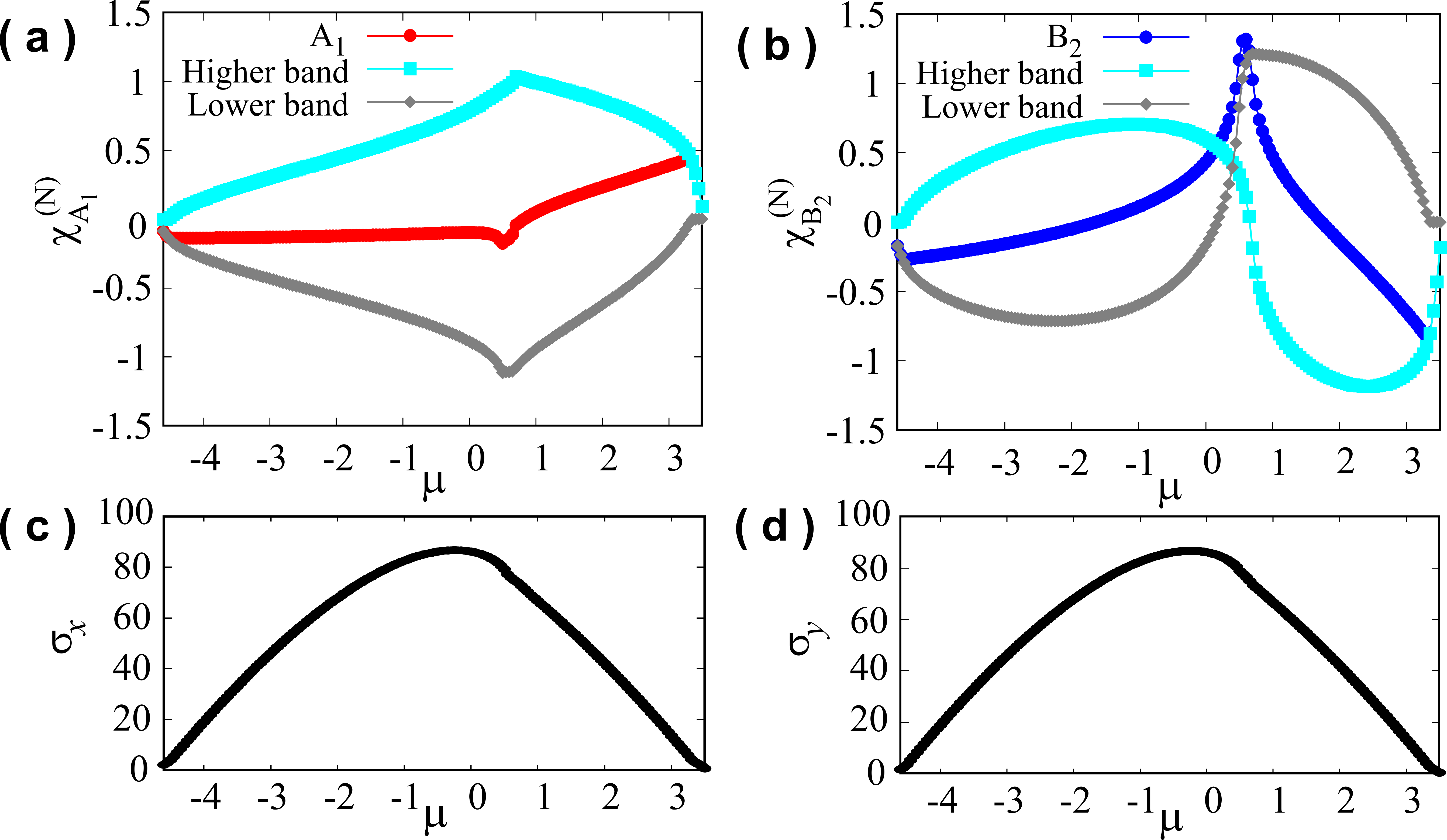}
  \caption{
Total susceptibility and band-resolved contributions for (a) $\chi^\textrm{(N)}_{\textrm{A}_1}$ and (b) $\chi^\textrm{(N)}_{\textrm{B}_2}$. (c) and (d) show the conductivity $\sigma_x$ and $\sigma_y$, respectively. 
We set the scattering rate $\delta = 0.01$. 
 }
\label{fig:chisig}
\end{figure}
As shown in Fig.~\ref{fig:chisig}, the two bands give almost opposite contributions, which cancel each other. In the low carrier density region, one band goes above or below the Fermi level, and the cancellation is suppressed. Therefore, the magnitude of $\chi^\textrm{(N)}_i$ increases as the Fermi level is approaching to the band edge. 
Similar discussions have been provided for other response functions in the Rashba system, such as the bulk rectification current~\cite{Ideue2017}. 
Furthermore, the electric conductivity $\sigma$ decreases there [Figs.~\ref{fig:chisig}(c) and \ref{fig:chisig}(d)], and thus, the MPE coefficients $d^\textrm{(N)}_i$ are rapidly enhanced. 
The peaks around $\mu = 0.6$ 
are attributed to the van Hove singularities at $\mu = 0.5,\,0.7$ for each band. Although they partially cancel out each other, the peak structure remains in $\chi^\textrm{(N)}_i$. 
We note that the peak structure of the MPE coefficients $d^\textrm{(N)}_i$ around $\mu=0.6$ is less pronounced than that in band edges because of the sizable conductivity.
These results corroborate that the MPE is determined by the Fermi surface effect~\cite{Watanabe2017} 
and thus distinct from the conventional PE.

\begin{figure}[t]
  \centering
  \includegraphics[width=70mm]{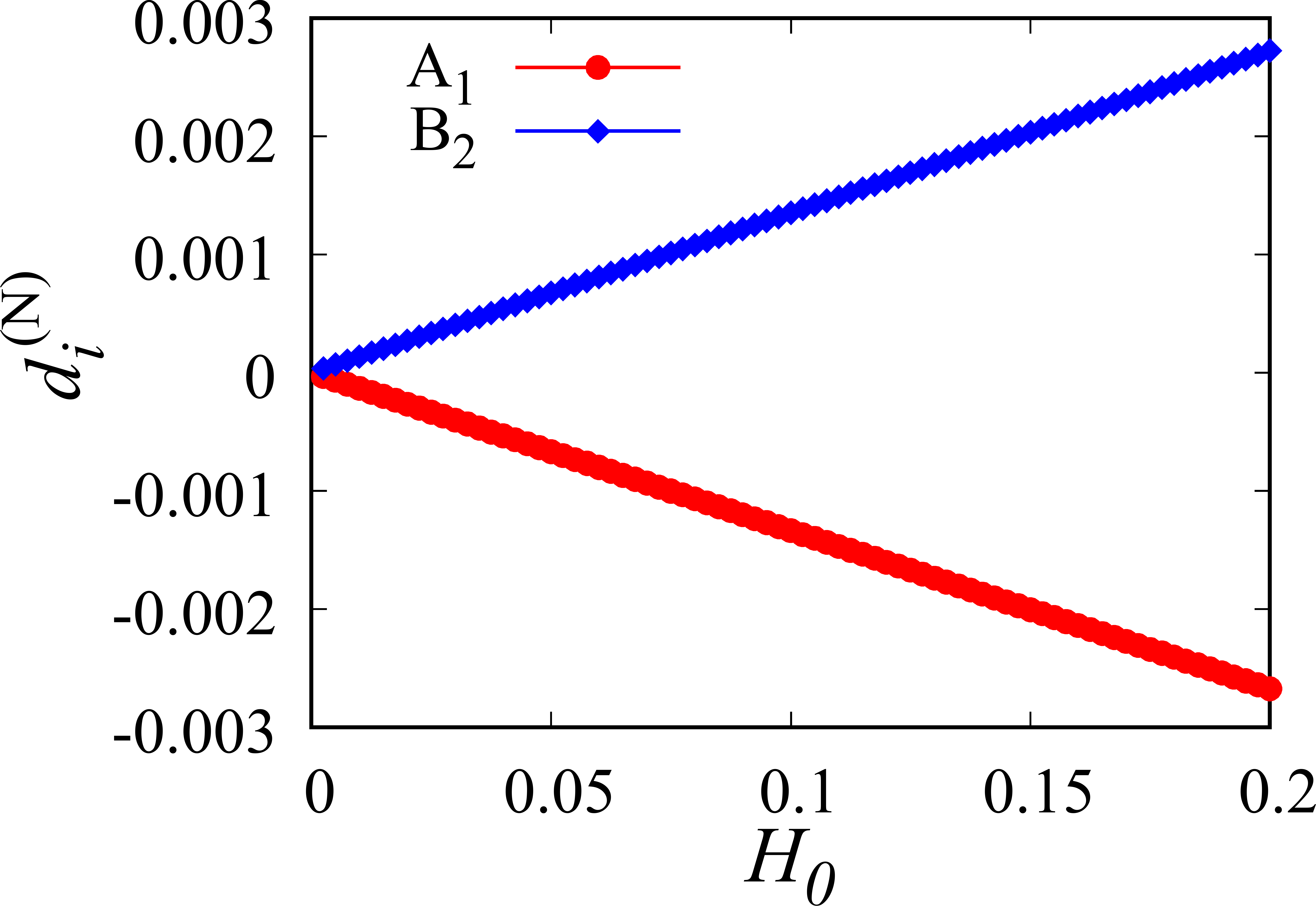}
  \caption{ 
 Magnetic field dependence of the MPE coefficients, $d^\textrm{(N)}_{\textrm{A}_1}$ and $d^\textrm{(N)}_{\textrm{B}_2}$. We set $\mu = -1.0$ to avoid influences of the characteristic band structures in the low carrier density region and near the van Hove singularities. 
 }
\label{fig:nmpeh}
\end{figure}

We also calculate the magnetic field dependence of $d^\textrm{(N)}_i$, and Fig.~\ref{fig:nmpeh} shows the result. We set $\mu = -1.0$ to avoid the effects of peculiar band structure around $\mu=0.6$ and the band edges.
The MPE coefficients are nearly proportional to the magnetic field and vanish at zero magnetic field. 
This linear relation indicates the controllability of the MPE, which is different from the previous theories~\cite{Watanabe2017,Watanabe2018} and experiments~\cite{Shiomi2019Aug,Shiomi2019Mar,Shiomi2020}. Previous studies worked on the parity-breaking antiferromagnet, which we can control by domain switching using the electric current~\cite{Zelezny2014,Wadley587,Watanabe2018b,Bodnar2018}. On the other hand, the MPE in noncentrosymmetric metals can also be controlled by the magnetic field.

\subsection{SCPE}
\begin{figure}[t]
  \centering
  \includegraphics[width=80mm]{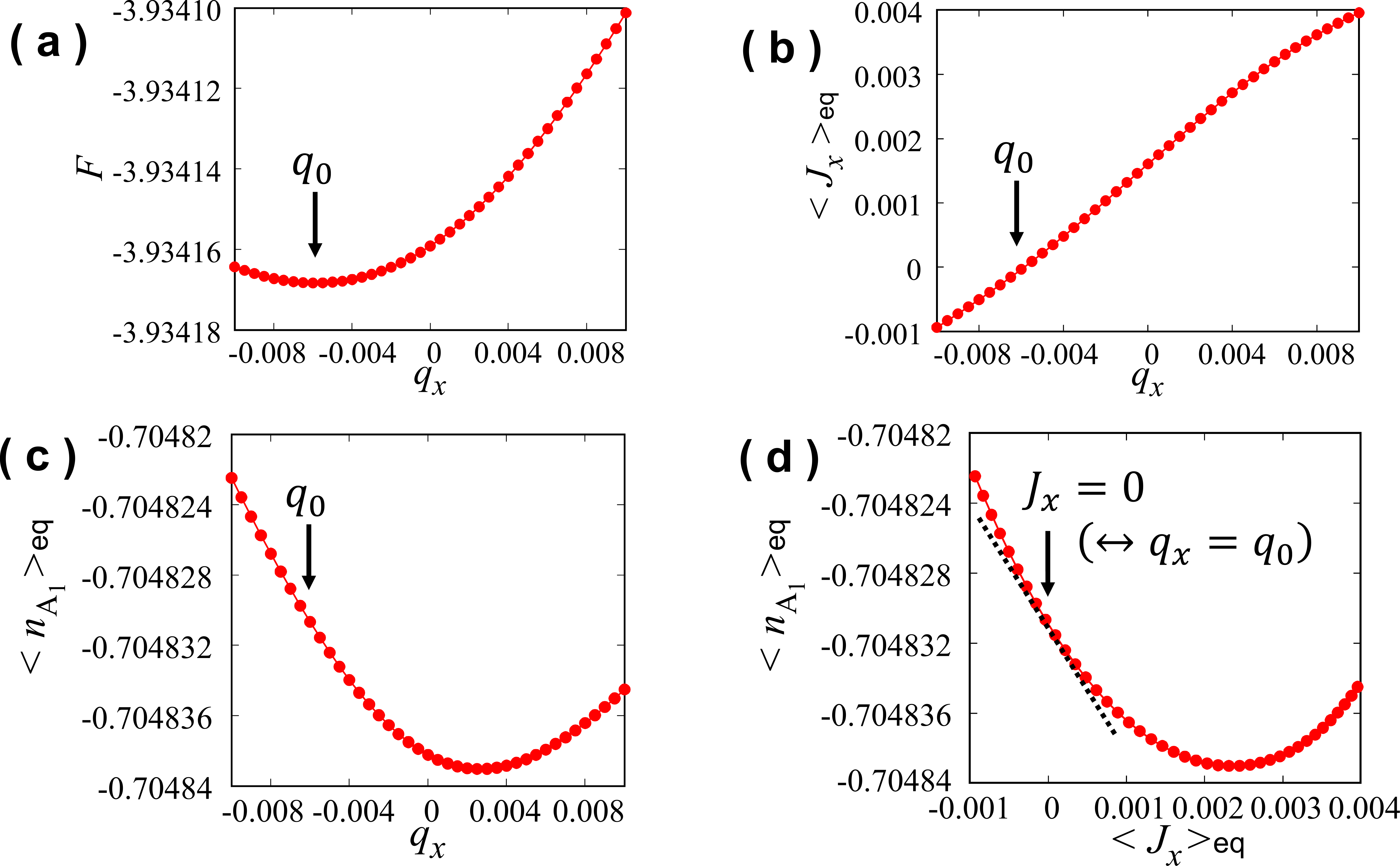}
  \caption{(a) Free energy $F$, (b) electric current $\ave{\hat{J}_x}_\text{eq}$, and (c) weighted density for the A$_1$ mode $\ave{\hat{n}_{\textrm{A}_1}}_\text{eq}$ as functions of $q_x$. Half of the momentum of Cooper pairs $q_0$ realizing the minimum free energy is illustrated by arrows. (d) $\ave{\hat{J}_x}_\text{eq}$ dependence of $\ave{\hat{n}_{\textrm{A}_1}}_\text{eq}$ obtained by combining the results of (b) and (c). We set $\mu = -1.0$ and $H_0 = 0.10$.
 }
\label{fig:scproc}
\end{figure}

Now let us discuss the SCPE.
Before showing the main results, we explain details of the calculation of SCPE coefficients by considering $d^\textrm{(SC)}_{\textrm{A}_1}$ as an example.
%
First, we determine the (half of) total momentum of Cooper pairs $q_0$ in the static  state. It is obtained so as to minimize the free energy given by
\begin{align}
F &= \ave{H}_\text{eq} - TS,\\
&= - \frac{T}{V} \sum_{\vk,\alpha} \log{\left(1+e^{-E_\alpha(\vk,\vq)/T}\right)}.
\label{FE}
\end{align}
To this end, we calculate the $q_x$ dependence of the free energy $F$ with the fixed chemical potential $\mu$ and magnetic field $H_0$ [see Fig.~\ref{fig:scproc} (a)]. Note that $F$ is symmetric with respect to $q_y$ and shows the minimum at $q_y = 0$ for any $\mu$ and $H_0$. Therefore, we find ${\bm q}=(q_0,0)$ on the $q_x$-axis, which characterizes the helical superconducting state with minimum free energy.
Next, we calculate the $q_x$ dependence of the electric current $\ave{\hat{J}_x}_\text{eq}$ and weighted density $\ave{\hat{n}_{\textrm{A}_1}}_\text{eq}$ around $q_x = q_0$  [Figs.~\ref{fig:scproc} (b) and (c)].
We can confirm the vanishing electric current $\ave{\hat{J}_x}_\text{eq} = 0$ in the static state with $q_x=q_0~(\ne 0)$. This is physically reasonable. Finally, combining these results, we obtain the $\ave{\hat{J}_x}_\text{eq}$ derivative of $\ave{\hat{n}_{\textrm{A}_1}}_\text{eq}$ around $q_0$ [Fig.~\ref{fig:scproc} (d)], which gives
the SCPE coefficient $d^\textrm{(SC)}_{\textrm{A}_1}$. 
In the numerical calculation, 
the derivative is evaluated by the difference around $q_0$ with a small momentum $q^\prime$
\begin{align}
d^\textrm{(SC)}_{\textrm{A}_1}
= \left. \frac{\delta \ave{\hat{n}_{\textrm{A}_1}}_\text{eq}}{\delta \ave{\hat{J}_x}_\text{eq}} \right|_{q_x = q_0}
= \frac{\ave{\hat{n}_{\textrm{A}_1}}_{\text{eq}, q_0 + q^\prime} - \ave{\hat{n}_{\textrm{A}_1}}_{\text{eq}, q_0 - q^\prime}}{\ave{\hat{J}_x}_{\text{eq}, q_0 + q^\prime} - \ave{\hat{J}_x}_{\text{eq}, q_0 - q^\prime}}.
\end{align}
We obtain the other SCPE coefficients $d^\textrm{(SC)}_{\textrm{B}_1}$ and $d^\textrm{(SC)}_{\textrm{B}_2}$ in the same way, 
while we examine the $q_y$ dependence instead of $q_x$ for the B$_2$ mode $d^\textrm{(SC)}_{\textrm{B}_2}$.
\begin{figure*}[!]
 \begin{center}
 \includegraphics[width=60mm]{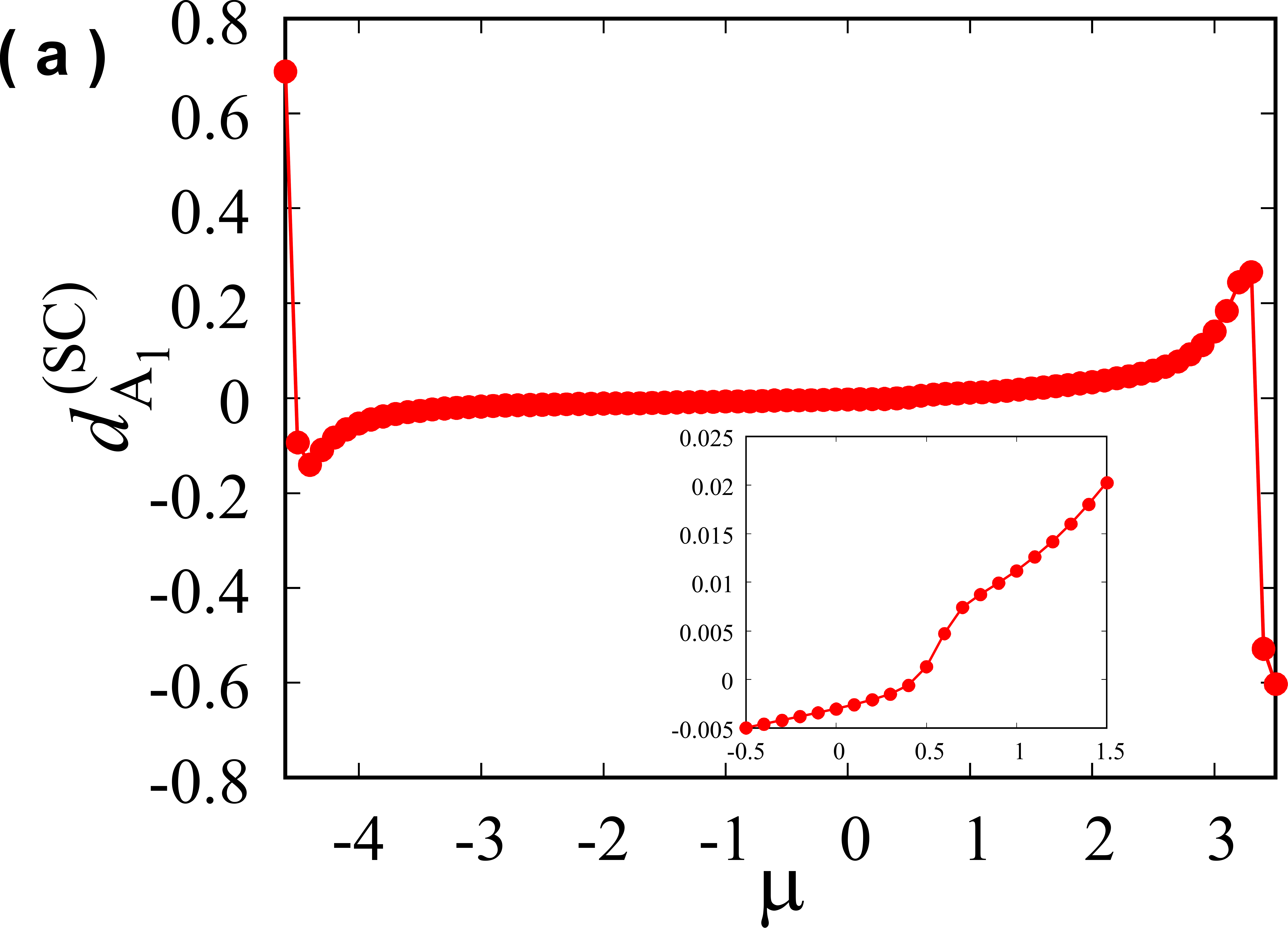}
 \includegraphics[width=60mm]{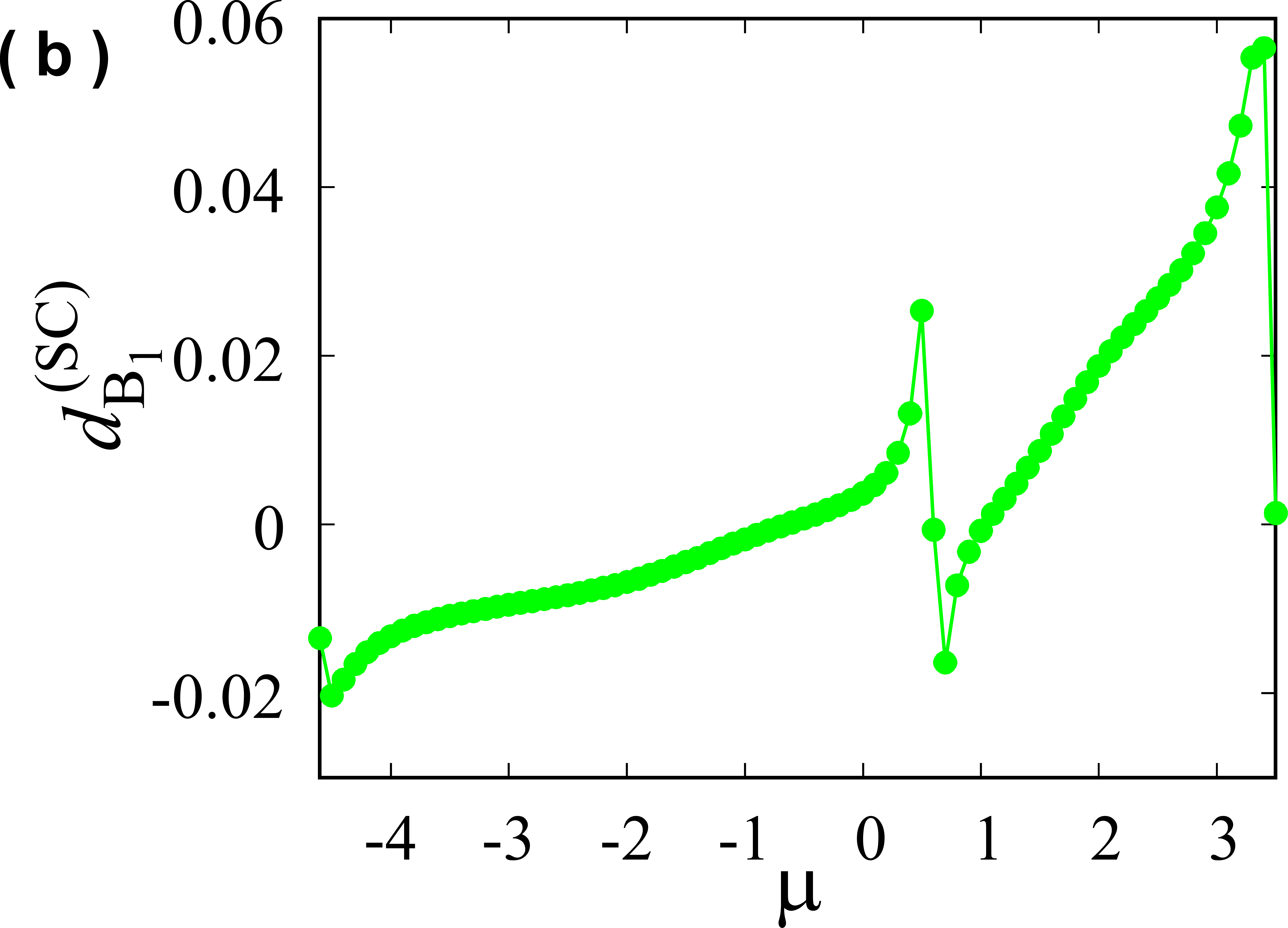}
 \includegraphics[width=60mm]{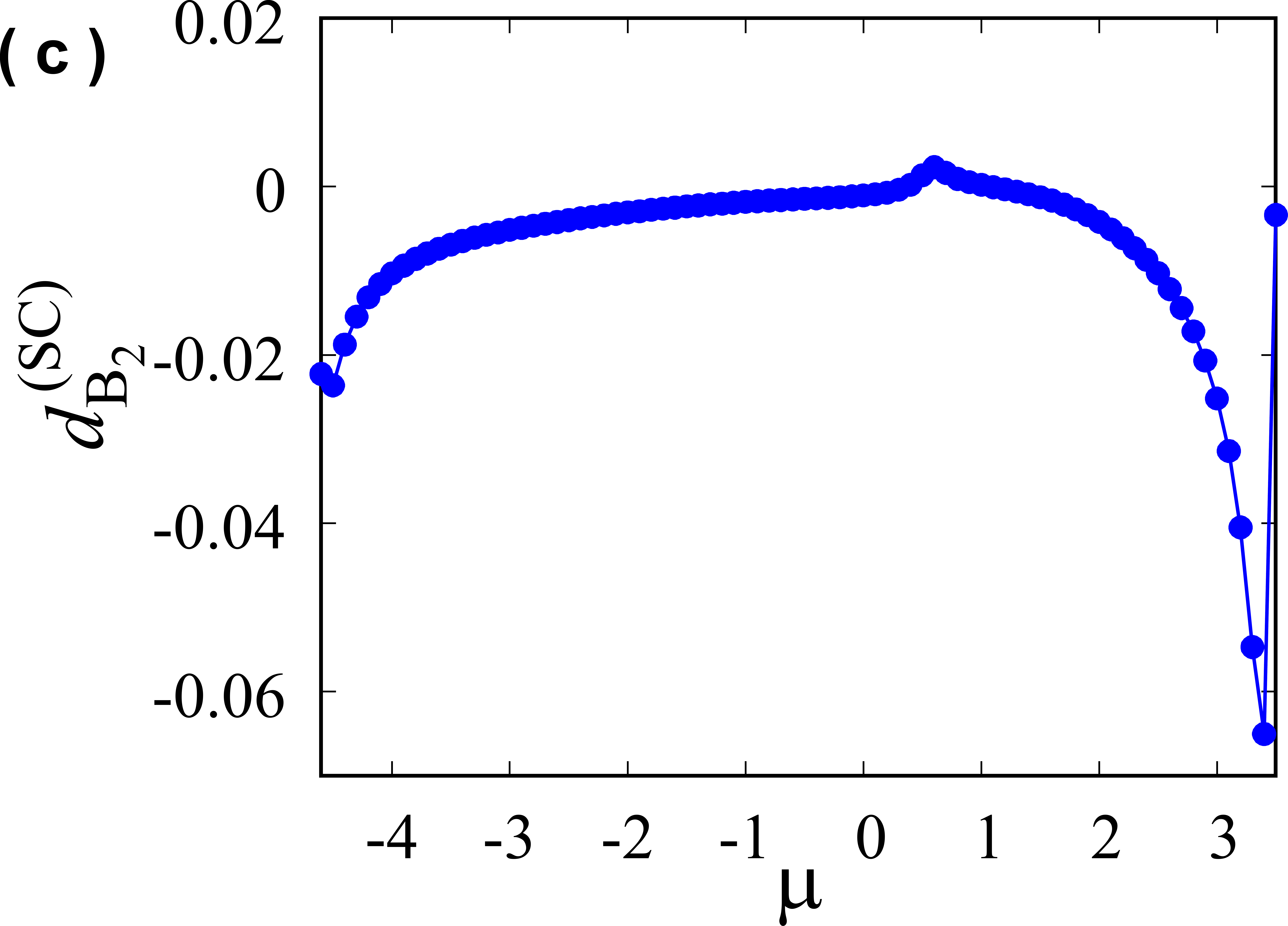}
 \includegraphics[width=60mm]{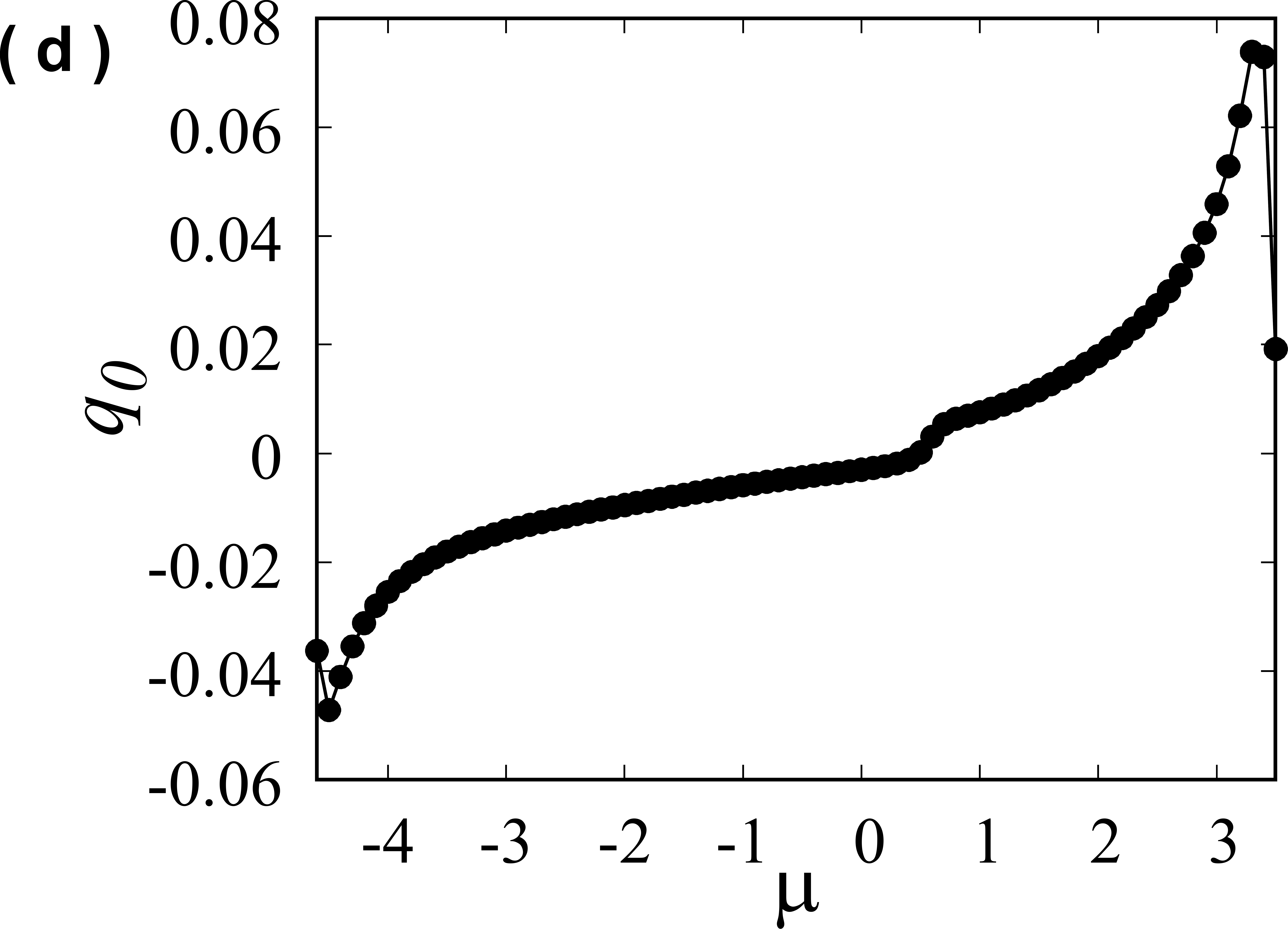}
  \caption{
SCPE coefficients (a) $d^\textrm{(SC)}_{\textrm{A}_1}$, (b) $d^\textrm{(SC)}_{\textrm{B}_1}$, and (c) $d^\textrm{(SC)}_{\textrm{B}_2}$. (d) Chemical potential dependence of $q_0$. Inset in (a) shows the region around $\mu = 0.6$.
}
 \label{fig:scpe}
 \end{center}
\end{figure*}

Then, we show the numerical results of the SCPE. The chemical potential dependence of the SCPE coefficients $d^\textrm{(SC)}_i$ and $q_0$ with the magnetic field $H_0 = 0.10$ is shown in Fig.~\ref{fig:scpe}. We obtain finite SCPE coefficients $d^\textrm{(SC)}_i$, and their magnitudes are comparable to the MPE coefficients $d^\textrm{(N)}_i$ in Fig.~\ref{fig:nmpe}. 
Because the SCPE and MPE coefficients are defined on an equal footing and the MPE was observed~\cite{Shiomi2019Mar,Shiomi2019Aug,Shiomi2020}, Fig.~\ref{fig:scpe} 
reveals the nonnegligible coupling of the supercurrent and lattice distortion.  

In Figs.~\ref{fig:nmpe} and \ref{fig:scpe}, we also find significant differences between the SCPE and MPE. Whereas the A$_1$ mode coefficient equals that of the B$_1$ mode in the MPE, they are different in the SCPE both qualitatively and quantitatively. 
The SCPE is smaller than the MPE for the B$_2$ mode, while the magnitude relation can be opposite for the A$_1$ mode. These results indicate that the SCPE and MPE are essentially different phenomena. This is reasonable because the source field is different between the SCPE and MPE (see Table.~\ref{table:PEs}). The supercurrent induces the SCPE without dissipation, although the dissipative current causes the MPE. Furthermore, the SCPE is not a Fermi surface effect because the excitation spectrum can be gapped in the superconducting state, while the MPE arises from the Fermi surface effect~\cite{Watanabe2017}. 

On the other hand, we also notice common features in $d^\textrm{(SC)}_i$, $d^\textrm{(N)}_i$, and $q_0$; all of them rapidly change in the low carrier density region and show a structure around $\mu = 0.6$. The chemical potential dependence of the normal MPE is explained by the band structure near the Fermi level as is evident from Eqs.~\eqref{chi2} and \eqref{sigma}. Thus, it is expected that the SCPE is also influenced by the Fermi surface in the normal state. 
\begin{figure}[b]
 \begin{center}
 \includegraphics[width=85mm]{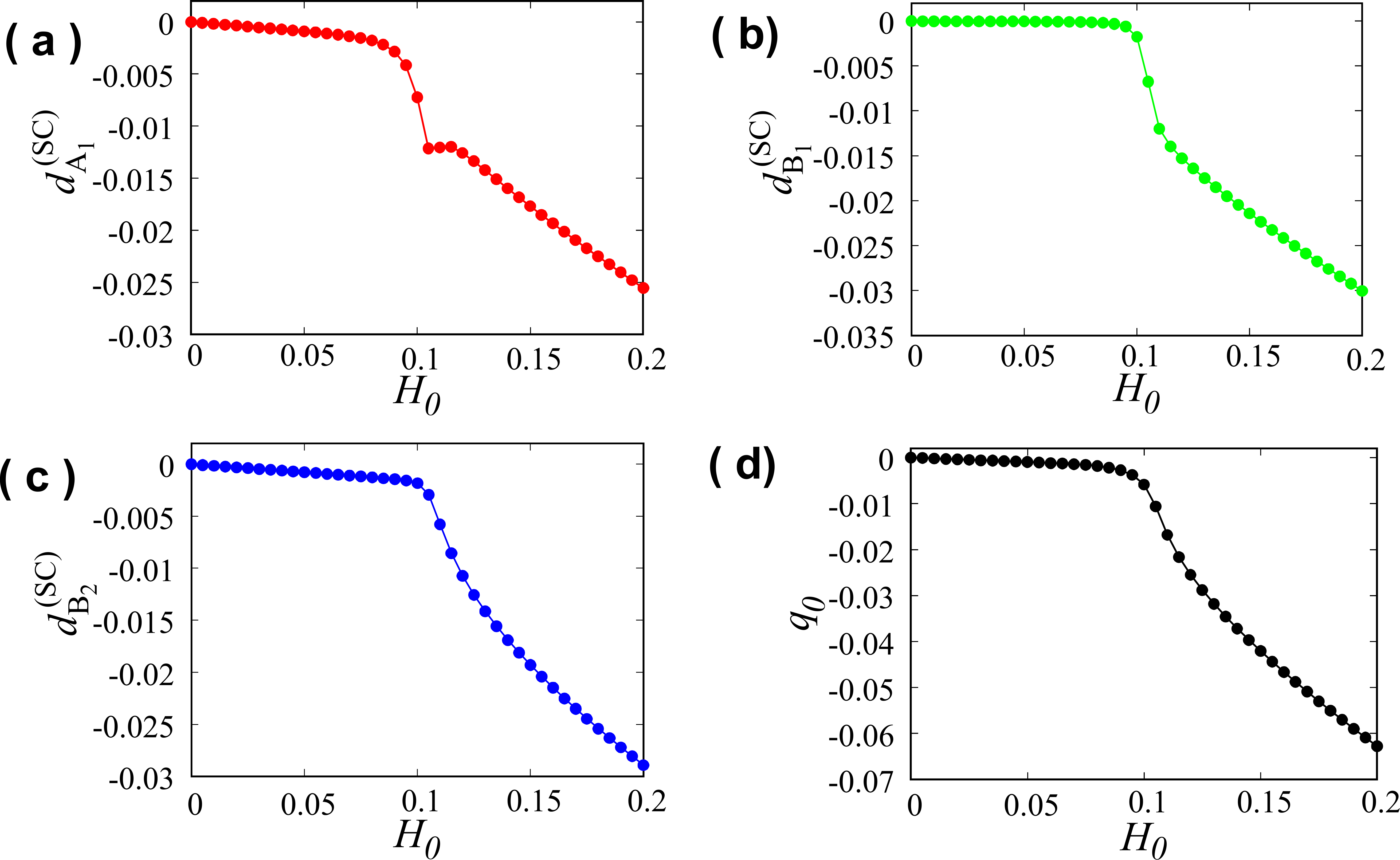}
  \caption{
Magnetic field dependence of the SCPE coefficients (a) $d^\textrm{(SC)}_{\textrm{A}_1}$, (b) $d^\textrm{(SC)}_{\textrm{B}_1}$, (c) $d^\textrm{(SC)}_{\textrm{B}_2}$, and (d) half of Cooper pairs' momentum $q_0$. We set $\mu = -1.0$ for the same reason as in Fig.~\ref{fig:nmpeh}. 
 }
 \label{fig:scpeh}
 \end{center}
\end{figure}

We understand the similar chemical potential dependence between the SCPE and MPE by considering the origin of Cooper pairs' momentum $2 q_0$. In the Rashba system, the degenerate bands are split by the spin-orbit coupling and shift in the opposite direction under an in-plane magnetic field~\cite{Bauer2012}. Because of the nonequivalence of the split bands, 
Cooper pairs have nonzero total momentum $2 q_0$ (helical superconducting state~\cite{Bauer2012}).
Thus, the behavior of $q_0$ is sensitive to the Fermi surface. Because the SCPE is induced by the supercurrent, it is most likely related to the Cooper pairs. Therefore, the SCPE coefficients $d^\textrm{(SC)}_i$ are indirectly affected by the Fermi surface through $q_0$ and then show the qualitatively similar behaviors to the MPE coefficients $d^\textrm{(N)}_i$. 

The magnetic field dependence of the SCPE further supports the essential role of Cooper pairs' momentum $2 q_0$ in the SCPE. As shown in Fig.~\ref{fig:scpeh}, the SCPE coefficients $d^\textrm{(SC)}_i$ do not show linear dependence on the magnetic field in contrast to the MPE coefficients $d^\textrm{(N)}_i$ in Fig.~\ref{fig:nmpeh}. The SCPE coefficients show abrupt change around $H_0 = 0.10$, and 
they follow the magnetic field dependence of $q_0$ [Fig.~\ref{fig:scpeh}(d)]. We see a remarkable similarity between the SCPE and Cooper pairs' momentum.
This indicates a close link between the SCPE and helical superconductivity. 
The nonlinear magnetic field dependence of $q_0$ has been explained by the crossover from the helical superconducting state to the Fulde-Ferrell state~\cite{Agterberg2007}. Thus, our results in Fig.~\ref{fig:scpeh} indicate that the SCPE could be useful not only to determine the symmetry of superconducting phases but also to probe the helical and Fulde-Ferrell states.

\section{SUMMARY AND DISCUSSION}
\label{sec:Summary}
We proposed the piezoelectric response in superconductors and named SCPE.  
We found that the SCPE is comparable to the MPE in magnitudes and clarified a close link with 
the helical superconductivity. 
 
Our results show that the MPE materials are good candidates for the SCPE materials. Therefore, it is expected that potential candidates hosting a sizable SCPE response can be found in materials with a large MPE response. Since the MPE and SCPE are enhanced by a strong antisymmetric spin-orbit coupling, some noncentrosymmetric superconductors may be favorable~\cite{Smidman2017}. Furthermore, the applied supercurrent flow can induce simultaneous breaking of the IS and TRS even in the absence of the intrinsic IS breaking~\cite{Nakamura2020,Yang2019,Vaswani2020}. Therefore, centrosymmetric superconductors are also candidates for the SCPE materials.

Search for the helical superconducting state with finite total momentum of Cooper pairs has been conducted in the noncentrosymmetric superconductors under a magnetic field~\cite{Bauer2012,Agterberg2007}. Indications for the helical superconductivity have been recently obtained in several superconductors via measurements of the upper critical field~\cite{Sekihara2013,Naritsuka2017-or,Naritsuka2021-ek} and nonreciprocal transport~\cite{Schumann2020-mw}, and direct observation in the superconducting state is awaited. Based on the finding of a close relation between the Cooper pairs' momentum and the SCPE, we proposed the probe of the helical superconducting state using the SCPE. 
In theoretical studies of the superconducting diode effect \cite{Yuan2021,Daido2021,he2021phenomenological}, which was recently discovered in experiments~\cite{Ando2020}, the importance of the helical superconductivity has also been pointed out, and the sign change in the nonreciprocal critical current is revealed to be a signature of the crossover in the helical superconducting state~\cite{Daido2021}. 
The SCPE is complementary to such phenomena and paves the way to detect the helical superconductivity by the linear response. 

The SCPE is expected to be useful for probing the symmetry breaking in superconductors because the IS and TRS breakings are required. 
Significantly, such spontaneous symmetry breaking is recently proposed in several superconductors. 
For examples, multiple superconducting phases with spontaneous IS and TRS breaking have been proposed in UTe$_2$ \cite{Ishizuka2021}, and TRS breaking has been reported in noncentrosymmetric superconductors, such as CaPtAs \cite{Shang2020} and so on \cite{Wysokinski2019}. In particular, discovery
of superconductivity in UTe$_2$ has stimulated vast studies for clarifying 
the spin-triplet superconducting state \cite{Ran2019}. However, recent observations of antiferromagnetic correlation~\cite{Thomas2020,Duan2020,knafo2021lowdimensional} also imply spin-singlet pairing. The possible coexistence of spin-triplet and spin-singlet Cooper pairs may lead to the spontaneous IS and TRS symmetry breaking~\cite{Ishizuka2021}, 
even though the crystal structure is centrosymmetric. It is urgent to determine the symmetry of multiple superconducting phases in UTe$_2$~\cite{Braithwaite2019,Lin2020,Aoki2020,Hayeseabb0272,Nakamine_2021}, and the SCPE may be helpful for solving the current issues.


\begin{acknowledgments}
We thank A.~Daido for fruitful comments.
This work was supported by JSPS KAKENHI (Grants No. JP18H05227, No. JP18H01178, and No. 20H05159) and SPIRITS 2020 of Kyoto University.
H.W. is a JSPS research fellow and supported by JSPS KAKENHI (Grant No.~18J23115 and No.~21J00453).
\end{acknowledgments}

\appendix
\section{MPE mode and relations between directions of \texorpdfstring{$\bm{J}$}{TEXT} and \texorpdfstring{$\bm{H}$}{TEXT}}
We rewrite the formula Eq.~\eqref{MPE} to
\begin{align}
s_{ij} = d^\prime_{ijkl} J_k H_l,
\label{MPE3}
\end{align}
by explicitly writing the external magnetic field $H_l$. 
Here, $d^\prime_{ijkl}$ should be invariant under the symmetry operations of the $C_{4v}$ point group characterizing the model Hamiltonian \eqref{HSC}.

In the two-dimensional system, we have three independent components for $d^\prime_{ijkl}$ in the C$_{4v}$ symmetry, and they are denoted as A$_1$, B$_1$, and B$_2$ modes. The strain tensors $s_{xx} + s_{yy}$, $s_{xx} - s_{yy}$, and $s_{xy}$ are involved with the A$_1$, B$_1$, and B$_2$ modes, respectively. The couplings of these modes and external magnetic fields are 
$J_x H_y - J_y H_x$, $J_x H_y + J_y H_x$, and $J_x H_x - J_y H_y$, respectively. 
For example, the allowed components for the A$_1$ mode
are denoted by $d^\prime_{xxxy}=-d^\prime_{xxyx}=d^\prime_{yyxy}=-d^\prime_{yyyx}$.

Following the above symmetry analysis of $d^\prime_{ijkl}$, we understand the relation between the applied electric current and induced strain. When we take $\bm{H} \parallel \hat{y}$ as in Fig.~\ref{fig:nem}, the A$_1$ and B$_1$ modes are induced by the current $\bm{J}$ perpendicular to the magnetic field $\bm{H}$, while the B$_2$ mode is induced by $\bm{J}$ parallel to $\bm{H}$. In contrast, when we set ${\bm H} \parallel [110]$, the A$_1$ and B$_2$ modes are induced when $\bm{J}$ is perpendicular to $\bm{H}$, and the B$_1$ mode is induced when $\bm{J}$ is parallel to $\bm{H}$. 

\section{Equivalence of the MPE coefficients $d^\textrm{(N)}_{\textrm{A}_1}$ and $d^\textrm{(N)}_{\textrm{B}_1}$}
We can diagonalize the normal state Hamiltonian \eqref{HN} using the normalized unitary matrix
\begin{align}  
& U(\vk) =
\frac{1}{\sqrt{2} \left| g_+ (\vk) \right|}
\left(
\begin{array}{cc}
 g_{+} (\vk) & - \left| g_+ (\vk) \right| \\
 \left| g_+ (\vk) \right| & g_{-} (\vk)
\end{array}
\right), 
\label{Umat} 
\end{align}
and the eigenvalues are given by
\begin{align}
E_n (\vk) 
= [U^\dag (\vk) H_{N} (\vk) U(\vk)]_{n n}
= \varepsilon (\vk) \pm \left| g_+ (\vk) \right|.
\label{En}
\end{align}
We then calculate the the band representation of the velocity operator $\tilde{v}_x(\vk)$ and obtain
\begin{align}
[\tilde{v}_x(\vk)]_{nn}
= \left[ U^\dag (\vk) \frac{\partial H_{N}(\vk)}{\partial k_x} U(\vk) \right]_{n n}
= \frac{\partial E_n(\vk)}{\partial k_x}.
\label{vn}
\end{align}
Therefore, Eq.~\eqref{chi2} is transformed into
\begin{align}
\chi^\textrm{(N)}_{\textrm{A}_1} 
&= \frac{- e}{\delta} \sum_{n} \int^{\pi}_{-\pi} \int^{\pi}_{-\pi}
D_{\textrm{A}_1}(\vk) \frac{\partial E_n(\vk)}{\partial k_x} \frac{\partial f(E_n)}{\partial E} 
\frac{dk_x}{2\pi} \frac{dk_y}{2\pi} \nonumber \\
&= \frac{- e}{\delta} \sum_{n} \int^{\pi}_{-\pi} \int^{\pi}_{-\pi}
D_{\textrm{A}_1}(\vk) \frac{\partial f(E_n(\vk))}{\partial k_x} 
\frac{dk_x}{2\pi} \frac{dk_y}{2\pi},
\label{chi3}
\end{align}
in the thermodynamic limit $V \rightarrow \infty$. 

Since both the A$_1$ and B$_1$ modes are induced by the electric current in the $x$-direction, the difference between $\chi^\textrm{(N)}_{\textrm{A}_1}$ and $\chi^\textrm{(N)}_{\textrm{B}_1}$ is calculated as follows,
\begin{align}
&\chi^\textrm{(N)}_{\textrm{A}_1} - \chi^\textrm{(N)}_{\textrm{B}_1} \nonumber \\
&= \frac{- e}{(2 \pi)^2 \delta} \sum_{n} \int^{\pi}_{-\pi} \int^{\pi}_{-\pi} \left( D_{\textrm{A}_1}(\vk) - D_{\textrm{B}_1}(\vk) \right) \frac{\partial f(E_n(\vk))}{\partial k_x} dk_x dk_y \nonumber \\
&= \frac{- 2e}{(2 \pi)^2 \delta} \sum_{n} \int^{\pi}_{-\pi} \int^{\pi}_{-\pi} \cos{k_y} \frac{\partial f(E_n(\vk))}{\partial k_x} dk_x dk_y \nonumber \\
&=\frac{- 2e}{(2 \pi)^2 \delta} \sum_{n} \int^{\pi}_{-\pi} \cos{k_y} \left[ f(E_n(\pi, k_y)) - f(E_n(-\pi, k_y)) \right] dk_y \nonumber \\
&= 0.
\end{align}
Thus, we have $\chi^\textrm{(N)}_{\textrm{A}_1} =  \chi^\textrm{(N)}_{\textrm{B}_1}$. 
The MPE coefficient $d^\textrm{(N)}_i$ is defined as Eq.~\eqref{NMPE}, and therefore, $d^\textrm{(N)}_{\textrm{A}_1}$ and  $d^\textrm{(N)}_{\textrm{B}_1}$ are also exactly the same.


\bibliography{article}

\end{document}